# Revealing Nanoscale Ni-Oxidation State Variations in Single-Crystal NMC811 *via* 2D and 3D Spectro-Ptychography

R.F. Ziesche[1,2,†,*], M.J. Johnson[3,†], I. Manke[1,*], J.H. Cruddos[3,4], A.V. Llewellyn[3,4,5], C. Tan[3,4], R. Jervis[3,4,5], P.R. Shearing[4,6], C. Rau[2], A.J.E. Rettie[3,4,5,*], S. Cipiccia[2,7], and D. Batey[2]

Enabling lithium (Li)-ion batteries with higher energy densities, longer cycle life, and lower costs will underpin the widespread electrification of the transportation and large-scale energy storage industries. Nickel (Ni)-rich layered oxide cathodes, such as $LiNi_xMn_yCo_zO_2$ (NMC, x > 0.8), have gained popularity due to their high specific capacities and lower cobalt content. However, the standard polycrystalline morphology suffers from accelerated degradation at voltages above 4.2 V versus graphite, due to its increased mechanical and chemical instability. Single-crystal NMC (SC-NMC) has emerged as a promising morphology for suppressing the mechanical instability by preventing intergranular cracking; however, robust methods of understanding its chemical degradation pathways are required.

We demonstrate how a high-throughput data collection strategy unlocks the ability to perform 2D and 3D ptychography in minutes, where it currently requires hours, and combine this with X-ray absorption near edge spectroscopy (XANES) to visualise the local Ni oxidation state behaviour in SC-NMC811, with nanometre-scale spatial resolution, and use this as a proxy for state-of-charge. By employing this technique at various stages during lifetime cycling to high voltages (>4.2 V), direct mapping of chemical degradation along the Li channels, identification of nucleation sites, and observation of Ni oxidation state heterogeneities across both electrode and particle scales can be achieved. We further correlate these heterogeneities with the formation and growth of the rocksalt phase and oxygen-induced planar gliding. This methodology will advance the fundamental understanding of how high-Ni layered oxide materials chemically degrade during high-voltage operation, guiding the design of more durable battery materials.

## Introduction

Lithium (Li)-ion batteries have become irreplaceable as modern energy storage systems, powering applications from portable electronics to electric vehicles and grid-level energy storage. The increasing demand for higher energy density, extended cycle life, and improved safety necessitates continuous advancements in battery materials and their characterisation. Among the various cathode materials used in commercial batteries, layered oxide materials like

[1] Helmholtz-Zentrum Berlin für Materialien und Energie (HZB), Hahn-Meitner-Platz 1, 14109 Berlin, Germany.
[2] Diamond Light Source, Harwell Campus, Didcot, OX11 0DE, UK.
[3] Electrochemical Innovation Lab, Department of Chemical Engineering, UCL, London, WC1E 6BT UK.
[4] The Faraday Institution, Quad One, Harwell Science and Innovation Campus, Didcot, OX11 0RA, U.K.
[5] Advanced Propulsion Lab, UCL East, UCL, London E15 2JE, U.K.
[6] The ZERO Institute, University of Oxford, Holywell House, Osney Mead, Oxford, OX2 0ES, UK.
[7] Department of Medical Physics and Biomedical Engineering, UCL, Gower St, London, WC1E 6BT, UK.
[†] These authors contributed equally to this work.
[*] Corresponding authors: Dr. Ralf F. Ziesche, ralf.ziesche@helmholtz-berlin.de; Dr. Ingo Manke, manke@helmholtz-berlin.de; and Dr. Alex Rettie, a.rettie@ucl.ac.uk

LiNi$_x$Mn$_y$Co$_z$O$_2$ (NMC) have gained popularity due to their high energy density, large voltage ranges and adaptability to various applications[1,2].

Their composition has evolved, transitioning from lower nickel (Ni)-content variants such as NMC111 (1:1:1 ratio of Ni:Mn:Co) to Ni-rich chemistries such as NMC811. This shift is driven by the need to maximise energy density and reduce reliance on cobalt, a scarce and costly resource that raises ethical concerns regarding its extraction[3–5]. The transition metals play various roles. Ni serves as the main electrochemically active species, contributing to the material's high capacity by facilitating Li-ion de-/intercalation by charge balancing through redox processes. Mn and Co enhance structural and thermal stability, with the latter also being redox active[4]. However, the increased Ni content of NMC811 introduces challenges such as heightened chemical reactivity and susceptibility to degradation[6]. Many degradation modes, such as electrolyte oxidation, oxygen release and *c*-lattice collapse, occur at a lower potential for NMCs with high Ni content[7–10]. While conventional polycrystalline NMC (PC-NMC) materials have been widely adopted, they often suffer simultaneously from structural and chemical degradation, which makes it difficult to isolate the effect of each on capacity fade. This is particularly pronounced in high-Ni variants, where issues such as intergranular cracking and surface layer reconstructions are exacerbated due to heightened reactivity and oxygen loss[11–13]. Recently, single-crystal NMC (SC-NMC) materials, which contain micron-sized primary particles, have been investigated. The absence of grain boundaries suppresses intergranular crack formation and enhances the mechanical stability[14–16]. However, these materials can still suffer from oxygen loss reactions, leading to surface layer reconstructions and inevitably capacity fade. Literature suggests that high-Ni NMCs can suffer from a rocksalt surface layer (RSL), which causes pinning of the crystal lattice, preventing lattice collapse and increasing charge transfer resistance, therefore, restricting further delithiation (see Fig. S1)[17–19]. The RSL is caused by highly reactive Ni$^{4+}$-ions on the surface being readily reduced to Ni$^{2+}$-ions, increasing the likelihood of Ni$^{2+}$ diffusing into free octahedral sites in the Li-ion layer.

Tools that can characterise these types of materials across multiple length-scales are key for understanding the complex relationships between the electrochemistry and performance loss. Conventional probes include small-angle X-ray/neutron scattering, which can provide information on the distribution of particle size and shape, and have been used to investigate the lithiation process in PC-NMC[20–23]. Complementary information is often provided by X-ray diffraction (XRD), which offers insights into lattice parameters, phase transitions, or strain in NMC during dis-/charge to understand rapid degradation[9,19]. X-ray absorption spectroscopy (XAS) reveals chemical information from the variation of the X-ray absorption coefficient with energy. Typically, X-ray absorption near-edge structure (XANES) is used to study changes of the oxidation state of a specific element and, therefore, can be used to study capacity fading during high C-rate cycling[24], upper cut-off voltages[25] leading to transition metal dissolution[26], or studying surface layer reactions[19]. Conventionally, these techniques offer a spatially averaged view of atomistic behaviour, typically with micron-scale resolution. More advanced techniques, such as XRD- and XANES-computed tomography (XRD-CT and XANES-CT, respectively), and Bragg coherent diffraction imaging (BCDI), leverage the benefits of microscopy combined with diffraction or spectroscopy, which can significantly improve the spatial resolution, albeit at the expense of analysis volume[27–29]. Higher resolution is accessible with electron and X-ray microscopy techniques such as transmission electron/X-ray microscopy (TEM/TXM), which have been used to study the evolution of surface layers on particles[30] or degradation of the cathode electrolyte interphase (CEI)[30]. The main drawback of TEM is the sample size and damage due to the low penetration power and absorption of electrons, and is therefore, often performed in 2D only. X-rays provide higher penetration through matter and enable operando 3D studies of commercial cells with μm resolution[31,32]. High resolution (on the order of tens of nm) is achievable with full-field TXM, which enables the investigation of the 3D morphology of active materials both *ex-situ*[33,34] and operando[35]. TXM and XANES have been successfully combined to investigate the charge distribution in PC-NMC[36–38], reversibility during dis-/charging [39] and protective coatings[40]. A disadvantage of TXM is the high X-ray dose delivered to the sample during the acquisition, which can cause local beam damage[41].

An alternative high-resolution X-ray imaging technique is X-ray ptychography, which is a scanning, coherent, small-angle scattering imaging method with nm resolution[42]. A ptychography dataset consists of a sequence of diffraction patterns obtained from overlapping regions of the sample illuminated with a coherent beam. From this data, the electron density and absorption properties are retrieved *via* iterative algorithms[43–47]. The technique is dose-efficient and has proven instrumental in studying battery morphology in 3D with high spatial resolution on secondary NMC particles using hard X-rays[48], and chemical oxidation states in 3D by comparing the measured absorption and X-ray energy of lithium iron phosphate (LFP) at two energies using soft X-rays[49]. When performed as an energy scan, ptychography provides a method for accessing XANES information using the retrieved absorption of the sample as a function of energy. This has been used to map the 2D chemical state of lithium nickel manganese oxide (LNMO)[50] and LFP[51]. Performing spectro-ptychography in 3D (3D-SP) requires a 4D scan (3 spatial dimensions and one energy dimension), which is very time-consuming. As a result, hard X-ray 3D-SP has to date only been demonstrated by Gao *et al.*,[28] following a 20-hour scan with sparse energy projections.

Here, we leverage the high-throughput data collection strategy of I13-1, the coherence imaging beamline at Diamond Light Source[45,52], to unlock the ability to perform 3D ptychography in minutes, where it currently requires hours. This is combined with X-ray absorption near-edge spectroscopy (XANES) to investigate the capacity fading mechanisms in SC-NMC811 by studying the local variation of Ni oxidation state in pristine and cycled samples, *via* hard X-ray far-field 2D-spectro-ptychography (2D-SP) and 3D-SP. 3D-SP reveals the local Ni oxidation state inside the SC-NMC811 particles, while 2D-SP produces maps across a larger sample size. We demonstrate, for the first time, the heterogeneity of intra- and interparticle Ni oxidation states and correlate this information with the pouch-cell level electrochemical performance. This understanding is critical to improving the lifetime performance of these materials by correlating degradation with chemical instability at the surface of the SCs during high-voltage cycling [38]. These insights can inform the design of more robust materials and strategies to mitigate the challenges of oxygen loss and surface phase transitions.

## Results

**Electrochemical characterisation.** Three commercially manufactured 200 mA.h SC-NMC811 vs. graphite pouch cells were used to investigate the long-term cycling behaviour of SC-NMC811. The morphology of these SC-NMC811 particles is shown in Fig. 1a, with an average crystal size between 1 and 3 μm. All long-term cycling was performed with constant current-constant voltage charging (CC-CV) to an upper cut-off voltage of 4.4 V for 200, 350 and 500 cycles, respectively. This was chosen to accelerate known degradation mechanisms that occur for high-Ni NMCs above 4.2 V[53–58]. There is more accessible lithium at these voltages, at which NMC811 exhibits additional structural phase transitions (Fig. 1b).

Fig. 1c shows the capacity retention vs. cycle number for each pouch cell, and three stages of degradation were identified. Stage I represents the steady cycle degradation observed when cycling to higher voltages than 4.2 V, with end-of-life (EoL) capacity (<80 %) being reached by approx. 200 cycles. When continuing beyond the EoL, stage II occurs, with an average capacity reduction of 50% over 150 cycles. Lastly, continuing into stage III, a capacity plateau occurred with only a further 5% capacity reduction over the next 150 cycles. To understand the source of this capacity loss, the relaxation voltages (RV) post-charge and discharge were investigated (Fig. 1d). Here, the RV at the top-of-charge was seen to steadily decrease from 4.30 V to 4.12 V during the duration of the testing. However, the RV at the bottom-of-discharge (BOD) was observed to drift significantly with age, most significantly in stage II, rising from 3.36 V to 3.65 V after 500 cycles. This behaviour can be indicative of difficulty in de-intercalating from graphite or intercalating into NMC. Additionally, this could be evidence of voltage slippage caused by the loss of accessible Li or a shift in the electrode alignment due to side reactions and gas formation.

Potentiostatic electrochemical impedance spectroscopy (PEIS) was performed every 50 cycles to monitor the change in resistance with ageing. Fig. 1e shows the Nyquist plot for the 500-cycle cell and fitting by the equivalent circuit model (ECM) shown. This ECM was informed *via* distribution of relaxation times analysis (see Fig. S2). From the fitting, we observe an increase in electrolyte resistance, which can indicate electrolyte depletion from fresh solid-electrolyte-interphase and CEI formation. Additionally, there is an initial increase in charge transfer resistance (CTR), followed by a subsequent decrease in CTR until 300 cycles, after which it begins to increase again. This decrease in CTR could be attributed to the BOD voltage slippage observed over the initial 350 cycles (Fig. 1d), which significantly contributes to the total cell resistance (Fig. 1f). Afterwards, RSL formation may become thick enough to increase the CTR and total cell resistance.

**3D-Spectro-Ptychography.** For the 3D analysis of Ni oxidation state changes, 3D-SP was performed on the pristine, 200- and 500-cycle SC-NMC811 materials. Fig. 2a-b show slices of the reconstructed attenuation coefficient and phase before and after the Ni K-edge for the pristine material. The slices display an agglomerate of SC-NMC811 crystals, which can be well distinguished from each other. This agglomeration is not typical of the SC architecture, but as also seen in Fig. 1a there is a small proportion of material that adopts this morphology. The attenuation increases significantly over the absorption edge, as indicated by the increase in grey-scale value, unlike the phase, which changes only slightly. Consequently, the absorption was used to determine the local oxidation states due to the better signal-to-noise ratio. The fitted oxidation states of the sample are displayed as a volume in Fig. 2c (with a quarter removed), showing a relatively homogeneous intra- and inter-particle distribution of Ni (majority < $Ni^{3+}$).

Fig. 3a shows an overview of the 200-cycle crystals in a volume representation (with a quarter removed), to reveal the interior of the crystals. Already, the crystals exhibit heterogeneity in the Ni oxidation state, with areas around $Ni^{2+}$ (blue) and others above $Ni^{3+}$ (yellow to red) indicating regions that are resistant to lithiation. Compared to pristine material, greater intra-particle heterogeneity is evident for the same state-of-charge (SOC), which could be evidence of RSL formation blocking the lithiation of interior channels. In certain crystals, clear slabs of oxidation states can be identified; crystals labelled with "b" to "d" are selected as examples. They are displayed separately as full volumes and slices through each plane in Fig. 3b-d. The areas of high Ni oxidation state (red) are segmented and isolated for each crystal in Fig. 3e-g.

From the distribution of these areas, two main routes of degradation can be inferred. Firstly, higher Ni oxidation states can be found near the surface of the crystal and secondly, through the interior of the crystal in the form of slabs. Surface degradation can be caused by the highly oxidative environment, which can induce RSL formation (see Fig. S1). This blocks the subsequent lithiation of Li layers and, therefore, we observe a retained high Ni oxidation state behind these layers. In the second case, the orientation of these areas seems to run parallel to the *c*-axis, forming slabs as shown in Fig. 3e-g. Due to the layered structure of NMC[59], the most likely route of degradation is perpendicular to the *c*-axis and therefore, the crystal orientation was estimated from these patterns. These slabs are thought to be caused by oxygen loss-induced planar gliding (see Fig. S3). The high operating voltage induces oxygen loss, which causes increased electrostatic repulsion between transition metal layers (TML) and, therefore, planar gliding (see Fig. S4). This gliding exposes fresh surfaces to RSL formation, which then extends towards the centre of the crystals. Although the resolution of this technique cannot resolve thin RSL at these structural defects, the high Ni oxidation state behind blocked channels can be used as an indicator for its presence.

After 500 cycles, the slab pattern within the crystals has developed and becomes even more evident (Fig. 4a-d). Specifically, a high concentration of $>Ni^{3.2+}$ (red) is observed close to the crystal core. However, there are regions with a lower oxidation state towards the surface of the crystals, which could be indicative of thick RSL growth. The extensive high voltage cycling means that more SCs experience oxygen loss-induced planar gliding of the TMLs (see Fig. S3 and S4). This increased fracturing is evident in the 2D projections as shown in Fig. 4e, where the crystal orientation is marked by blue arrows. These results highlight the intra-particle Ni oxidation state heterogeneities that SC-NMC811 experiences when cycled to high operating voltages. The effects of oxygen loss and RSL formation on capacity fade are inferred from the increased interior and slab-like oxidation state behaviour observed. The increasing intra-particle heterogeneities with cycle number can be correlated with the capacity fade and highlight the importance of passivating oxygen loss and RSL formation in these materials.

**2D-Spectro-Ptychography.** To verify that the heterogeneities observed in the 3D-SP experiments were representative of the entire electrode, a larger sample size was investigated using 2D-SP. Fig. 5a-b show a single slice of the reconstructed phase and the inverse-logarithm of the modulus (an absorption equivalent) from the pristine SC-NMC811, showing a cluster of SCs. The Ni oxidation state was then inferred over every pixel from the Ni K-edge shift at the full width half maximum (FWHM)[59]. Fig. 5c-f shows representative crystals from the 2D oxidation state maps of the pristine, 200, 350 and 500 cycled crystals, respectively. The overall colour change of the maps from blue to red can infer the crystal SOC due to an increase of Ni oxidation state from $Ni^{2+}$ to $>Ni^{3+}$ as presented in the histogram in Fig. 5g. As the cycle number increases the distribution of Ni oxidation states across the 2D maps broadens, representing a growing inter-particle heterogeneity across the electrode with ageing.

The pristine crystals exhibited the narrowest distribution of Ni, with an average of around $Ni^{2.3+}$, whereas the average from the 200-cycle sample was closer to the $Ni^{3+}$ state. However, the distribution of the oxidation states is wider from $Ni^{2.3+}$ to $Ni^{3.5+}$ and shows higher oxidation states (red) exist towards the edges of crystals (Fig. 5d). Furthermore, higher oxidation states were observed in stripe patterns across some crystals. These stripe patterns are consistent with the slab-like heterogeneity observed in the 3D-SP, which we correlate with oxygen loss-induced planar gliding, and is also visible in SEM images (see Fig. S3 and S4)[59,60]. The distribution of Ni oxidation states along these defects suggests that planar gliding can cause heterogeneous relithiation of Li-ion layers.

The average oxidation state of the 350-cycle sample was $Ni^{3.1+}$, with the inter-particle heterogeneity broadening further (Fig. 5g). For example, there are crystals similar in oxidation state to the 200-cycle materials and others existing in a higher Ni oxidation state (see Fig. S8). The key difference being that the higher Ni oxidation states have spread from the crystal edges towards the centre (Fig. 5e). The higher proportion of high oxidation state material correlates with the pronounced capacity loss of 40% and indicates a reduced SOC window being utilised in the SC-NMC811 (Fig. 1c). After 500 cycles the overall Ni oxidation state increases only slightly in comparison to the 350-cycle material, e.g., see Fig. 5e-f. This observation is consistent with the plateauing nature of the capacity fade at this stage. The 2D oxidation state maps (see Fig. S9 and S10) show a large portion of crystals with an oxidation state >$Ni^{3+}$ (red) and a few crystals more consistent with the 200-cycle material. This could indicate that a reduced number of crystals are partially electrochemically active and contribute towards the diminished capacity. However, it is worth noting that the presence of crystals with lower oxidation states could be caused by a poor electrical connection due to gas formation or large potential drops across the electrode, meaning that isolated crystals may remain inactive and closer to their pristine state.

2D-SP reveals local Ni oxidation state heterogeneities within individual SC-NMC811 particles, enabling the investigation of large sample areas to interrogate inter-particle heterogeneities within an electrode. However, the clustering of crystals, their orientation with respect to the beam and the superposition of oxidation states over the crystal(s) make it impossible to study fine variations in oxidation state through the 3D volume. Highlighting the importance of performing complementary 3D-SP and 2D-SP experiments to understand the intra- and inter-particle heterogeneities that can exist, respectively.

## Conclusions

Spectro-ptychography (SP) was employed to reveal intra- and inter-particle Ni oxidation state heterogeneities that occur in SC-NMC811 at various stages of cycle life. We observe that high Ni oxidation states (> $Ni^{3+}$) occur in slab patterns from the surface through to the core of crystals, following planar gliding defects. Although the resolution of this technique cannot resolve the nm-thick rocksalt layers, the high Ni oxidation states at blocked Li-ion channels were assigned to the presence of these layers along the planar gliding regions. 2D-SP probed the heterogeneity at the particle and electrode level, showing that the distribution of Ni oxidation states broadened with cycle number, while the average Ni oxidation state increased. These findings are consistent with progressive oxygen loss and the formation of a rock salt layer, which inherently demonstrates their impact on capacity fade. These results highlight the significance of surface layer modifications to SC chemistries that prevent oxygen release and inhibit the formation of rock salt surface layers, thereby promoting prolonged capacity retention and cycle life.

2D- and 3D-SP have proven to be valuable tools for studying degradation mechanisms in SC Li-ion cathode materials due to the high spatial resolution of the oxidation state of the active Ni species. The method is widely applicable to all electrochemical systems for tracking SOC or degradation pathways. In addition, it is not restricted to crystalline structures, allowing the study of nanocrystalline, amorphous and liquid materials. The use of soft/hard X-rays allows chemical changes to be studied in 3D, whereas soft X-rays are limited to 2D due to depth penetration. In future work, SP will be developed into an operando/in-situ technique, allowing for the study of diffusion pathways through crystals/electrodes during operation. This is supported by complementary method developments such as multi-beam[61], broadband[62,63], and sparse-ptychography[64–66]. All improvements promise to reduce the measurement time of SP to a few minutes or hours for 2D or 3D, respectively. Additionally, the simultaneous collection of diffraction data alongside SP would allow for the correlation of chemical and crystal structure information.

Figure legends

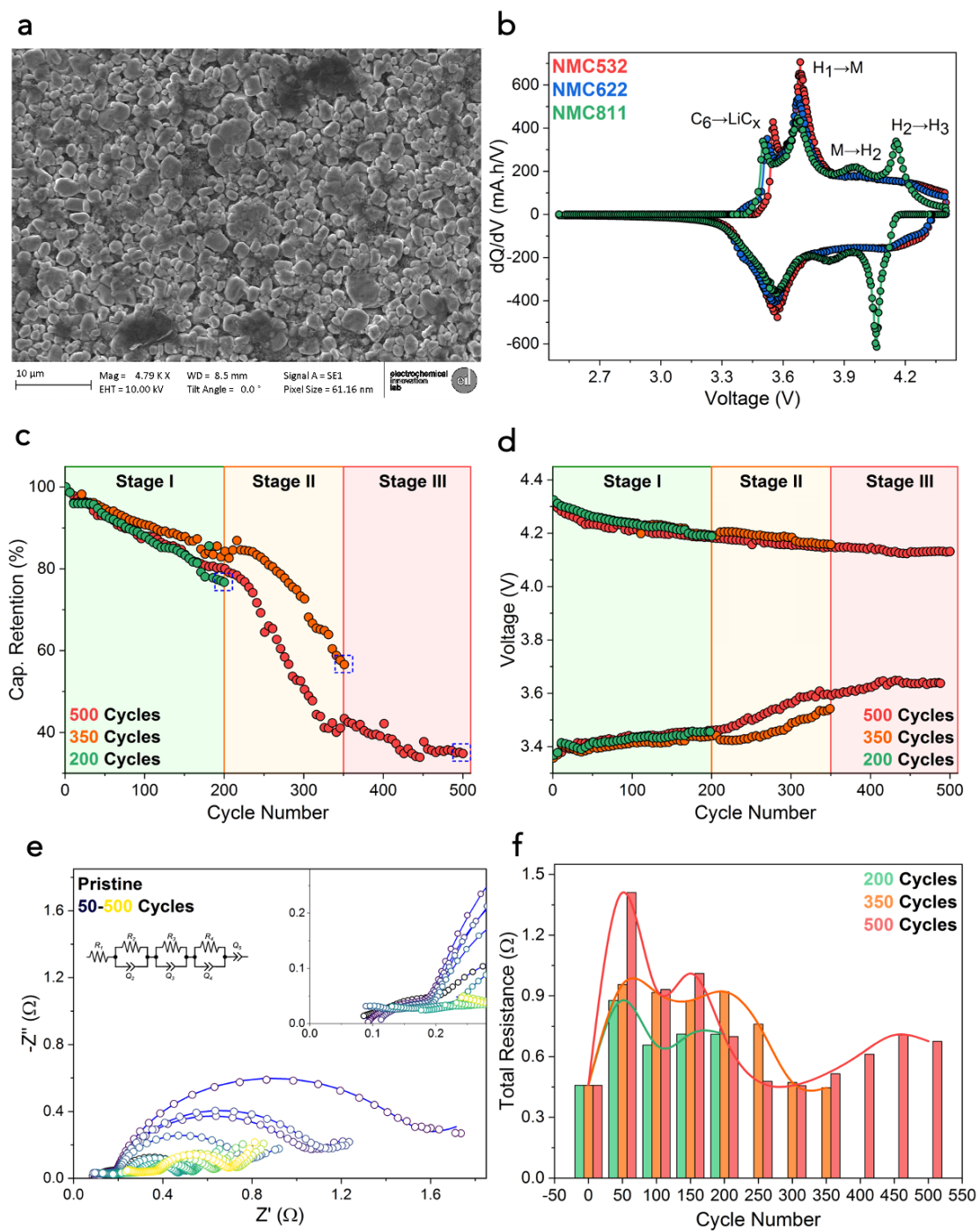

Fig. 1: | Electrochemical characterisation of SC-NMC811 pouch cells. **a,** Scanning electron microscopy (SEM) image of pristine SC-NMC811 electrode, against a 10 μm scalebar. **b,** dQ/dV vs. voltage for SC-NMC532, SC-NMC622 and SC-NMC811 pouch cells, showcasing the difference in phase transitions. **c,** Capacity retention vs. cycle number for three SC-NMC811 cells cycled to 200, 350 and 500 cycles. Shaded regions indicate the three stages of degradation. **d**, Open circuit voltage (OCV) approximation vs. cycle number for the three cells, based on the final voltage in the relaxation after the CC-CV charge (top) and CC discharge (bottom). **e,** Nyquist plot for the 500 cycled cell, with EIS collected every 50 cycles at the bottom-of-discharge during the lifetime cycling. Open circles are the experimental datapoints, and the red line is the fitting by the equivalent circuit model (ECM) shown in the plot. **f,** The total resistance vs. cycle number for all three cells, calculated from the ECM fitting.

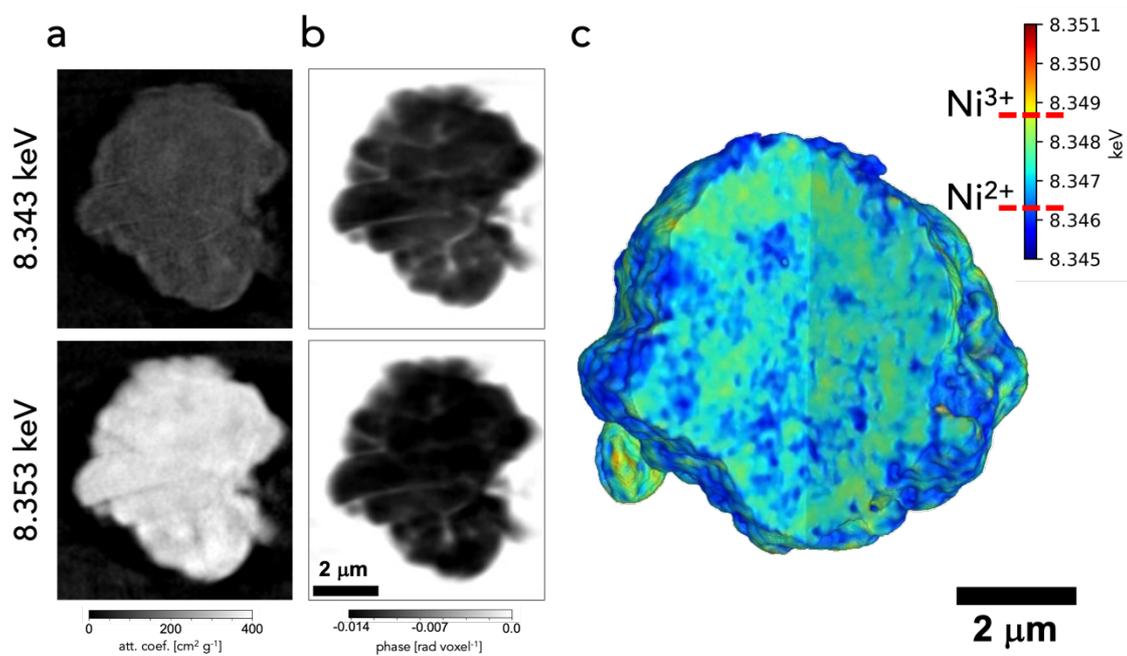

Fig. 2 | 3D spectral ptychography on pristine SC-NMC811. **a,** Vertical orthogonal slices of the attenuation coefficient and **b,** phase before and after the Ni K-edge. **c,** 3D rendering of the XANES-fit. All scale bars are 2 μm.

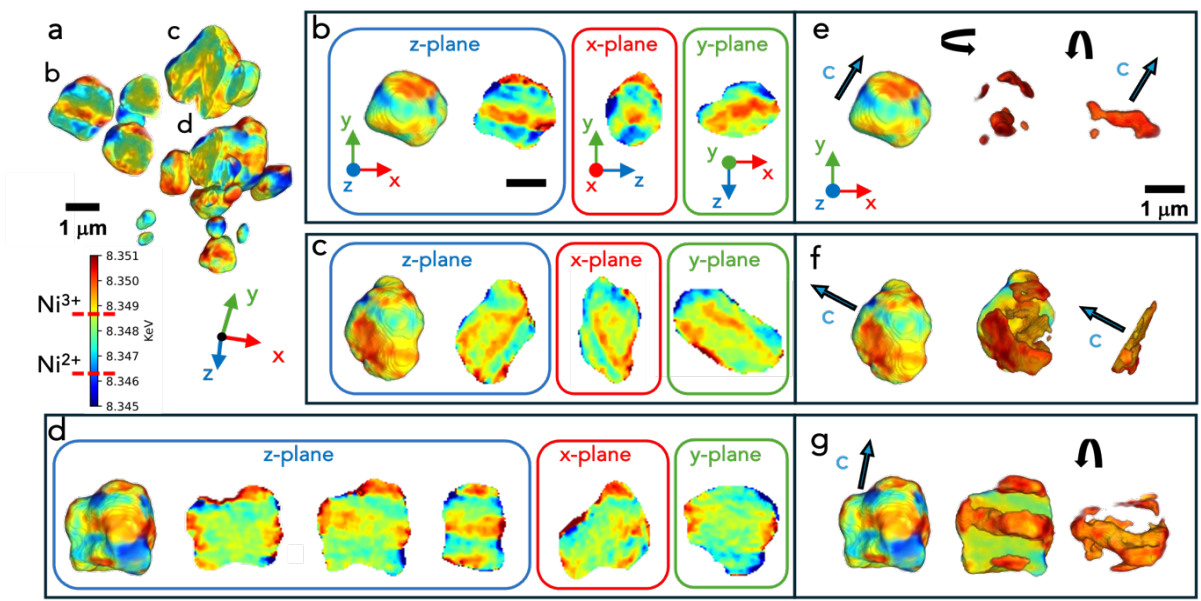

**Fig. 3 | 3D spectral ptychography on 200-cycle SC-NMC811. a,** 3D visualisation of XANES-fit showing variation of the Ni oxidation state from $Ni^{2+}$ to >$Ni^{3+}$. **b-d,** 3D visualisation and orthogonal slices of selected crystals in **a,** showing crystal ageing inside along each plane. **e-g** shows 3D rendering of the crystals from **b, c, d,** respectively and the segmentation of strong degraded areas of >$Ni^{3+}$. From the layered structure of NMC and the crystal shape, the slab-wise heterogeneity is assumed to run along the crystal *c*-plane as marked by blue arrows. All scale bars are 1 μm.

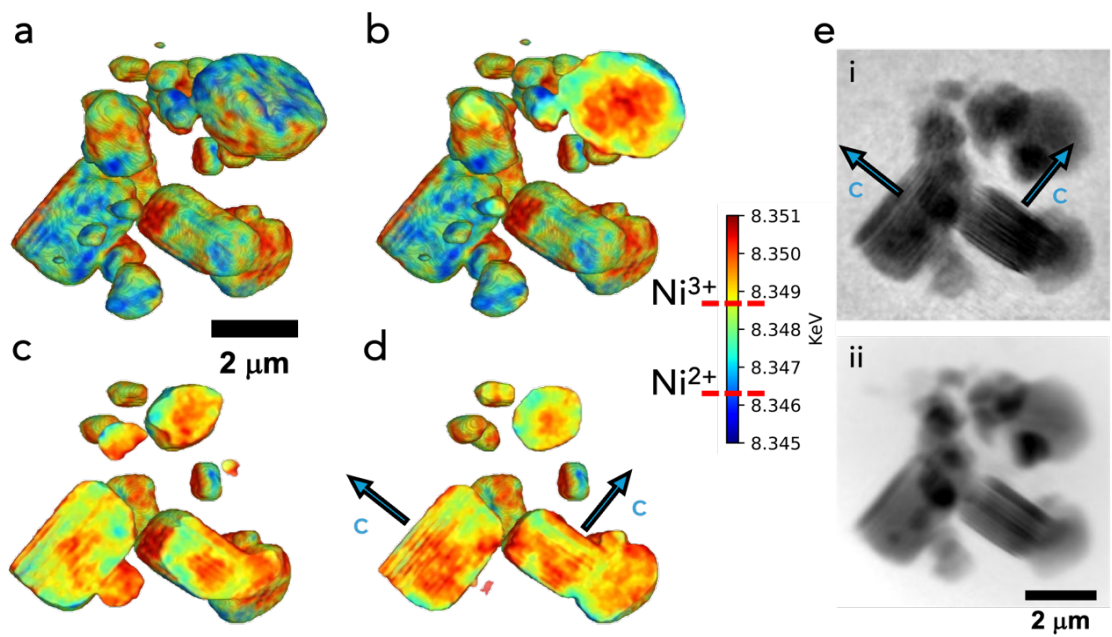

**Fig. 4 | 3D spectral ptychography on 500-cycle SC-NMC811. a-d,** 3D rendering of XANES-fitted crystals with progressing segmentation through the crystal to show internal Ni-oxidation state mapping, showing the propagation of >$Ni^{3+}$ oxidation state from the surface into the crystal centre. **e.i,** Phase and **e.ii,** modulus projections showing the layered delamination of crystals along the crystal *c*-plane. Scale bars are 2 μm.

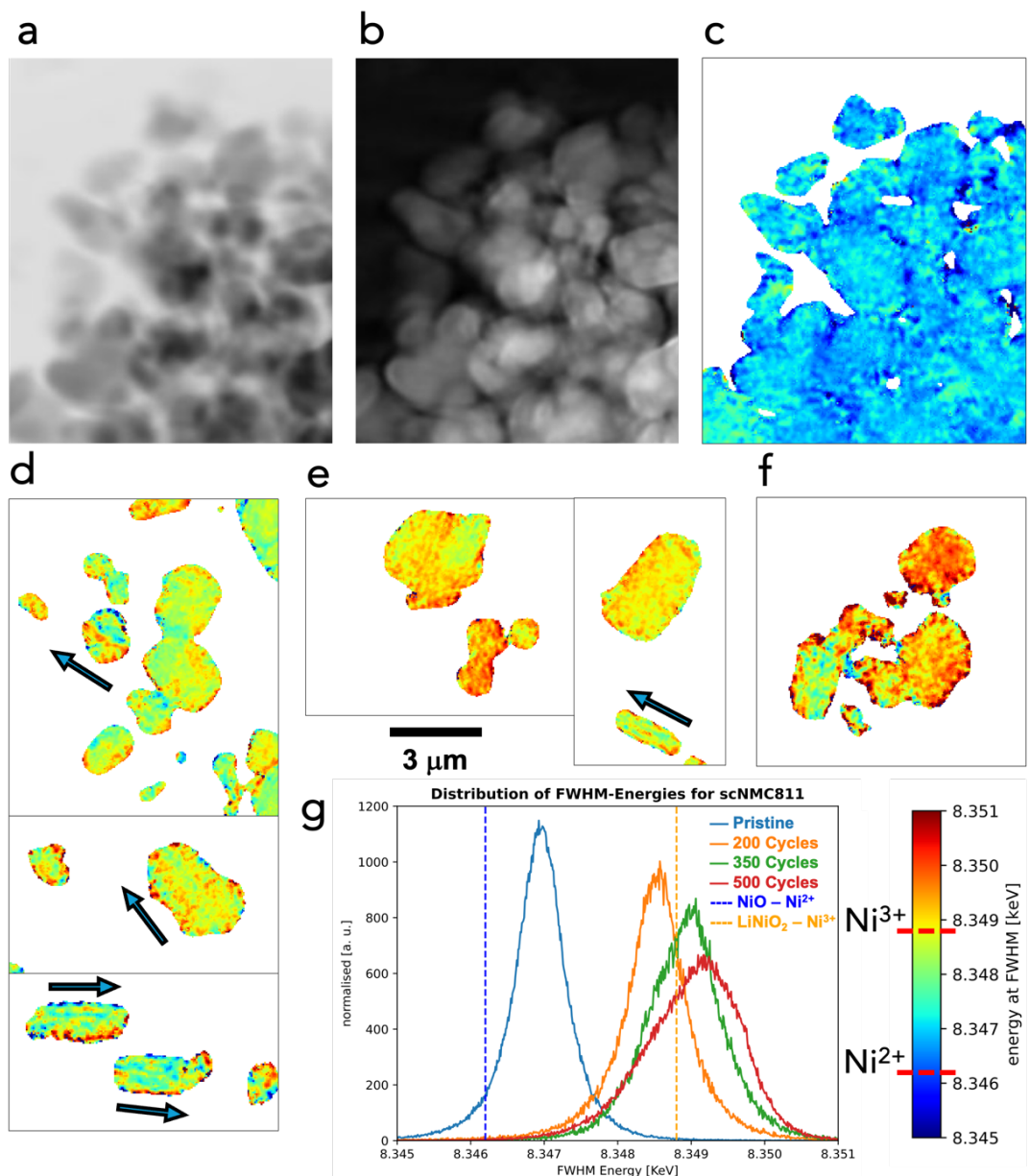

**Fig. 5 | 2D spectral ptychography on pristine and aged SC-NMC811. a** Phase and **b** inverse-log of the modulus of pristine SC-NMC811 and **c,** XANES-fit of the Ni K-edge of the same section. **d-f,** XANES-fit of a selection of individual SC-NMC811 crystals after 200, 350 and 500 cycles, respectively. All selected from 40 × 40 μm² scan areas shown in the supplementary information (see Fig. S5-S10). **g,** Histograms of the observed Ni oxidation states in each sample area taken from Fig. S5-S10. Dashed lines represent powder standards for $Ni^{2+}$ (NiO) and $Ni^{3+}$ ($LiNiO_2$) collected. Blue arrows indicate areas that follow the stripe pattern consistent with planar gliding. **a-f,** Scale bar is 3 μm.

## Methods

**Li-ion Pouch Cells.** The commercially manufactured cells used in these experiments were 200 mA.h nominal capacity pouch cells (Li-FUN Technology Corporation, Ltd, China). These cells were assembled using single-crystal Ni-rich ($LiNi_{0.8}Mn_{0.1}Co_{0.1}O_2$) cathodes and artificial graphite anodes. These cells are manufactured and formed by Li-FUN Technology Corporation, and therefore, electrolyte formulation and formation protocols cannot be provided here due to the company's intellectual property.

**Electrochemistry.** The cells were cycled using a Biologic BCS-805 battery cycler with 150 mA channels (Biologic, France). All cells performed C/2 (100 mA) constant current-constant voltage cycling (CC-CV) charging to 4.40 V with a C/20 (10 mA) current cut-off. Followed by CC discharge at C/2 (100 mA) until 2.70 V. After each charge or discharge step, the cells were held at open circuit voltage (OCV) for five minutes. Potentiostatic electrochemical impedance spectroscopy (PEIS) was performed using a 10 mV voltage perturbation over the frequency range of 10 kHz to 10 mHz, at the bottom-of-discharge. This cycling profile was repeated for 200, 350 and 500 cycles for the respective cells with PEIS collected every 50 cycles in all cases. All electrochemical testing was conducted at room temperature and in compression jigs (see Fig. S11).

**Scanning Electron Microscopy.** The scanning electron microscopy (SEM) data were collected using a Zeiss EVO MA10 microscope (Carl Zeiss AG, Germany) by loading samples onto adhesive carbon tape and exposing the sample to an electron beam operating at an accelerating voltage of 10 kV.

**Harvesting and Mounting of Single Crystals.** All SC materials were harvested from the commercially manufactured pouch cells (see Fig. S11). Pristine crystals were removed from dry electrode sheets using a surgical scalpel, inside an argon-filled glovebox. The 200-, 350- and 500-cycled crystals were obtained by opening discharged cells within an argon-filled glovebox, followed by washing the cathode sheets with anhydrous dimethyl carbonate (DMC, Sigma Aldrich, >99% purity) to remove any electrolyte salts. These sheets are then allowed to dry for 24 hours before removing the SC powders with a surgical scalpel. For 2D imaging, the SCs are then glued to strips of Kapton tape using UV-glue and then mounted on an aluminium frame with four imaging positions for the pristine, 200, 350 and 500 cycled samples, respectively (see Fig. S12). For 3D imaging, the samples were mounted onto Kapton loops (20 μm dual thickness MicroMounts, MiTeGen) using UV-glue, visualising the crystals with a Keyence VHX-7000 optical microscope (see Fig. S13).

**Synchrotron/Beamline.** All synchrotron experiments were performed at the I13-1 beamline at DLS – Diamond Light Source, Harwell Science and Innovation Campus, Didcot, United Kingdom.

*Ex-situ* 2D/3D Spectro-Ptychography Collection.

For both experiments, a Fresnel zone plate with a diameter of 400 μm and outermost zone width of 200 nm was used to focus the coherent X-ray beam and generate a beam spot size of approx. 2 μm at the sample placed just downstream of the focal plane of the lens. The diffraction patterns were recorded using an EIGER 500k photon counting detector (silicon substrate and 75 μm pixel size, from DECTRIS, Baden-Dättwil, Switzerland), 5 m downstream from the sample.

Each of the 2D samples was scanned over an area of approximately 40 × 40 μm$^2$ with a horizontal and vertical step size of 0.2 μm and 0.4 μm, respectively, in a 200 × 100-step grid. The X-ray energy was scanned across the Ni K-edge between 8.325 to 8.370 keV with 1 eV steps, and an additional three energies pre-edge (8.27, 8.285 and 8.30 keV) and two energies post-edge (8.40 and 8.42 keV) for the normalisation of the XANES data. The scanning imaging was performed in flight scan mode [45], with continuous sample movement in the horizontal direction, and an exposure time of 3 ms per scanning point (333 Hz). The data collection took approximately 60 s per energy step.

For the 3D-SP, 241 projections were taken over an angular range of 180°. Each projection was scanned with a step size of 0.3 µm in both the horizontal and vertical directions on a 50 × 50 step grid, resulting in a field of view of approximately 15 × 15 µm$^2$. The scan was performed in flight mode, resulting in ~ 6.75 s per projection and < 2700 s per CT. For all the samples imaged in 3D, the X-ray energy was scanned across the Ni K-edge between 8.341 to 8.360 keV with a 1 eV step size and the same pre- and post-edge energies (8.27, 8.285, 8.30, 8.40, and 8.42 keV).

**XANES Ptychography Reconstruction.**

For the 2D ptychography reconstruction of the phase and modulus (transmission), the implementation of the ePIE algorithm in the ptychography reconstruction code PtyREX[67] developed in-house at Diamond Light Source, was used. The retrieved projections were realigned and processed to produce a 3D tomographic reconstruction *via* an iterative reconstruction and reprojection algorithm, based on work by Gursoy[68]. For 3D reconstruction, the simultaneous iterative reconstruction technique (SIRT) algorithm was employed with 200 iterations. For reprojection, forward projection as implemented in the ASTRA toolbox [69,70] was utilised within a Python script. Following this, the energy-dependent 2D and 3D phase and modulus reconstructions were rescaled to a pixel/voxel size of 40 nm and registered using a phase cross-correlation algorithm[71].

**Fitting of Ni oxidation state.**

Fitting of the 2D and 3D data was done pixel- or voxel-wise on the modulus reconstructions. The 2D projections were converted to an integrated attenuation coefficient map by calculating the inverse-log of the modulus. Normalisation of the XANES data was done by using the pre- and post-edge. The Ni K-edge was then fitted by superimposing three Gaussian functions over the absorption edge. The oxidation states were determined by the shift of the absorption edge at the FWHM, calculated from the 1$^{st}$ main peak of the Ni K-edge (see Fig. S14). For the calibration of the oxidation state of nickel, Ni-metal, NiO and LiNiO$_2$ were measured as references for Ni, Ni$^{2+}$ and Ni$^{3+}$, respectively (see Fig. S15).

## Methods references

## Data and code availability

Correspondence and requests for materials should be addressed to the corresponding author: Dr. Ralf F Ziesche.

## Acknowledgements

We gratefully acknowledge Diamond Light Source for the provision of beamtime (MG30867-1, MG35179-1). PRS acknowledges the Royal Academy of Engineering for the Chair in Emerging Technologies (CiET1718/59). MJJ acknowledges HORIBA-MIRA, UCL and EPSRC (EP/R513143/1) for a CASE studentship that supported these experiments. We would like to thank the German ministry (BMWI, BMWK, BMWE) for their support (project HiBrain 30ETE039G) and the German DFG (Project DAPHNE, no. 460248799). The authors would like to thank Peng Li for beamline support at I13-1. The Faraday Institution's Degradation project, funded by grants FIRG0024, FIRG060, and FIRG082, and LISTAR project, funded by grants FIRG058 and FIRG083 supported this work.

## Author contributions

RFZ and MJJ both contributed equally as co-first authors. RFZ, MJJ, SC and DB led the investigation and contributed to the main project idea and experimental conceptualisation. RFZ, MJJ, SC, DB, CT and CR conceived the experiments. RFZ, MJJ, SC, DB, JHC and AVL performed the experiments. RFZ, MJJ, SC, DB, and JHC performed the data analysis. DB and CR lead work at Diamond Light Source as the scientists in charge of I13-1. RJ, PRS and AJER led work from the Electrochemical Innovation and Advanced Propulsion Labs (UCL) and sourced funding to support this work. AJER contributed to data interpretation, funding acquisition and supervision. IM contributed to the data interpretation, conceptual ideas, fundraising and supervision.

## Competing interest declaration

There are no competing interests known to the authors to declare.

## Additional information

Supplementary information is available for this paper.

## Corresponding author

*Dr. Ralf F. Ziesche, ralf.ziesche@helmholtz-berlin.de

*Dr. Ingo Manke, manke@helmholtz-berlin.de

*Dr. Alex Rettie, a.rettie@ucl.ac.uk

# Supplementary Information

# Revealing Nanoscale Ni-Oxidation State Variations in Single-Crystal NMC811 *via* 2D and 3D Spectro-Ptychography

R.F. Ziesche[1,2,†,*], M.J. Johnson[3,†], I. Manke[1,*], J.H. Cruddos[3,4], A.V. Llewellyn[3,4,5], C. Tan[3,4], R. Jervis[3,4,5], P.R. Shearing[4,6], C. Rau[2], A.J.E. Rettie[3,4,5,*], S. Cipiccia[2,7], and D. Batey[2]

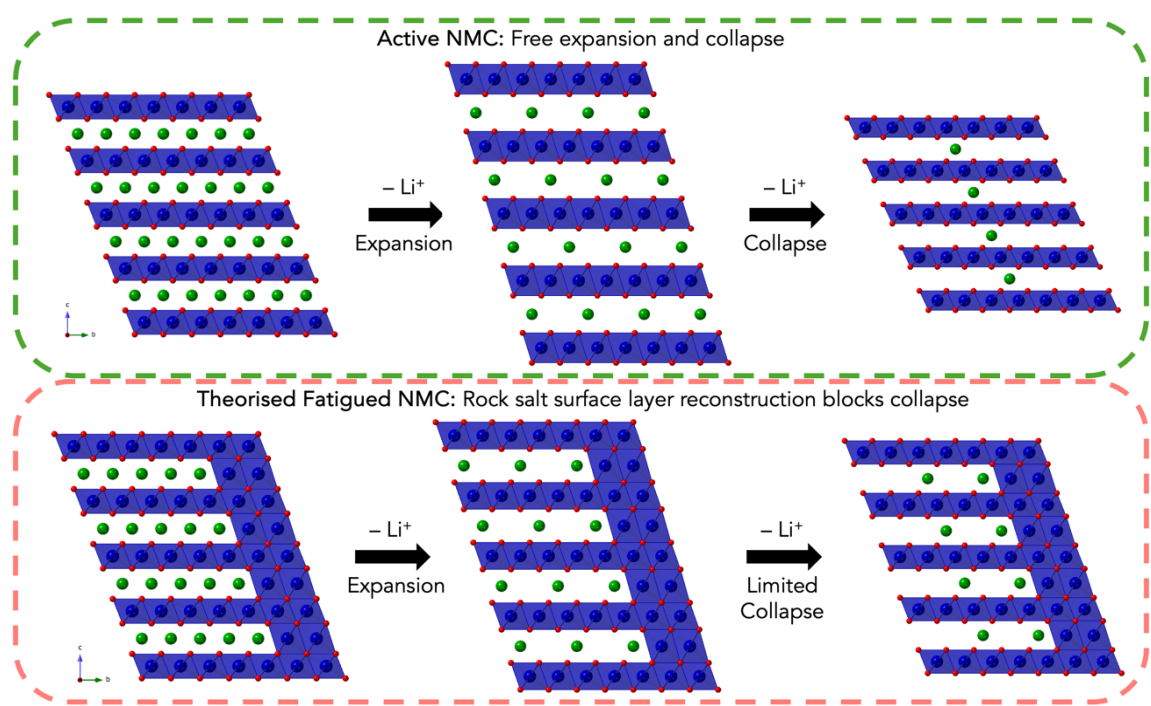

Fig. S1: a Schematic representation of the rock-salt surface layer transformation pinning the crystal lattice open, preventing the *c*-lattice collapse and further delithiation [1].

[1] Helmholtz-Zentrum Berlin für Materialien und Energie (HZB), Hahn-Meitner-Platz 1, 14109 Berlin, Germany.
[2] Diamond Light Source, Harwell Campus, Didcot, OX11 0DE, UK.
[3] Electrochemical Innovation Lab, Department of Chemical Engineering, UCL, London, WC1E 6BT UK.
[4] The Faraday Institution, Quad One, Harwell Science and Innovation Campus, Didcot, OX11 0RA, U.K.
[5] Advanced Propulsion Lab, UCL East, UCL, London E15 2JE, U.K.
[6] The ZERO Institute, University of Oxford, Holywell House, Osney Mead, Oxford, OX2 0ES, UK.
[7] Department of Medical Physics and Biomedical Engineering, UCL, Gower St, London, WC1E 6BT, UK.
† These authors contributed equally to this work.
* Corresponding authors: Dr. Ralf Ziesche, ralf.ziesche@helmholtz-berlin.de; Dr. Ingo Manke, manke@helmholtz-berlin.de; and Dr. Alex Rettie, a.rettie@ucl.ac.uk

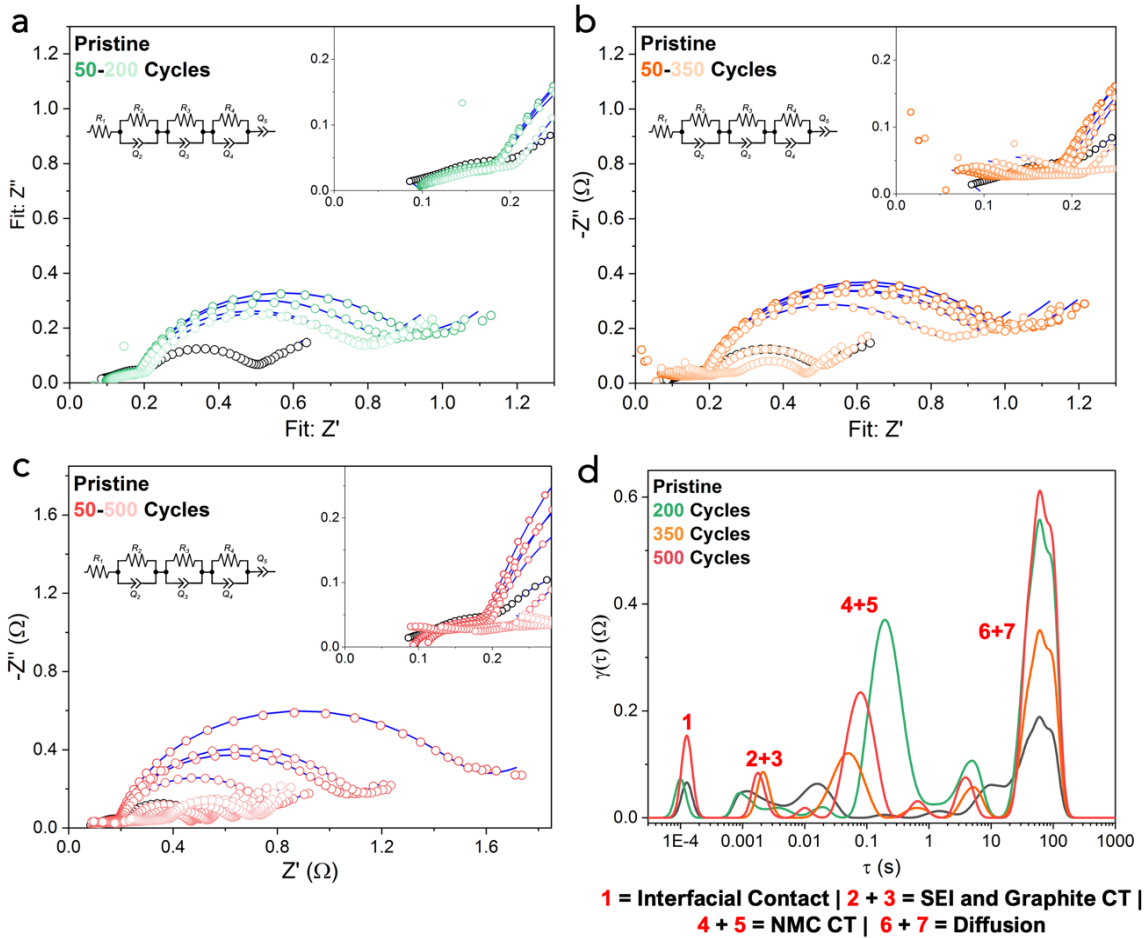

Fig. S2: a-c, Nyquist plots for the pristine, 200-, 350- and 500-cycle cells, respectively. The open circles represent experimental data points, and the line represents the fit by the equivalent circuit model (shown in the inset). d, Distribution of relaxation times (DRT) analysis for pristine, 200-, 350- and 500-cycle cells.

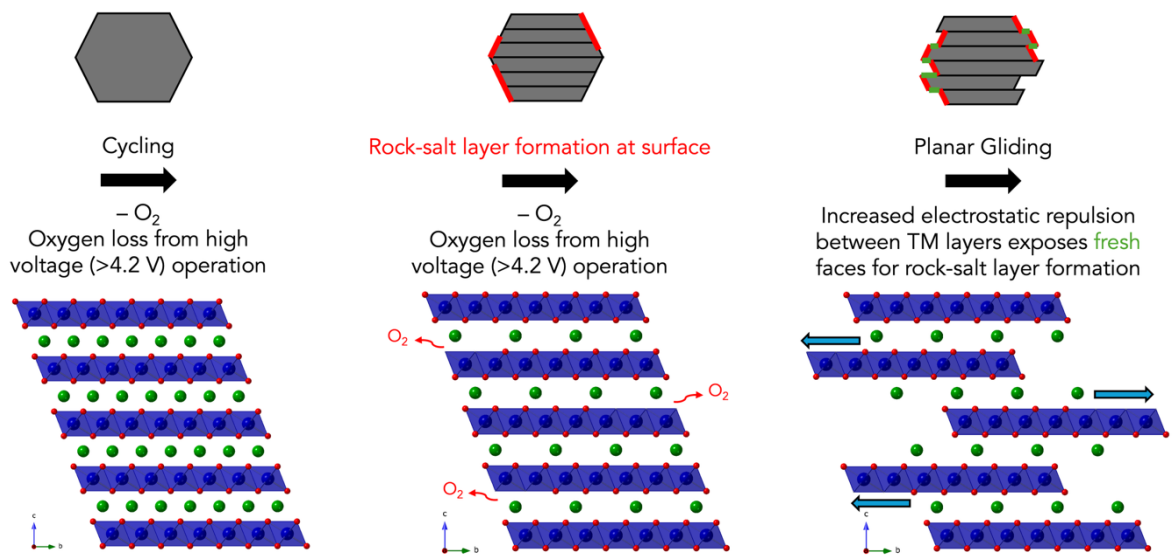

**Fig. S3:** A Schematic representation of the mechanism of oxygen loss-induced planar gliding in NMC due to electrostatic repulsion of the transition metal-ion layers. This in turn increases the surfaces that are readily available for rock-salt layer formation.

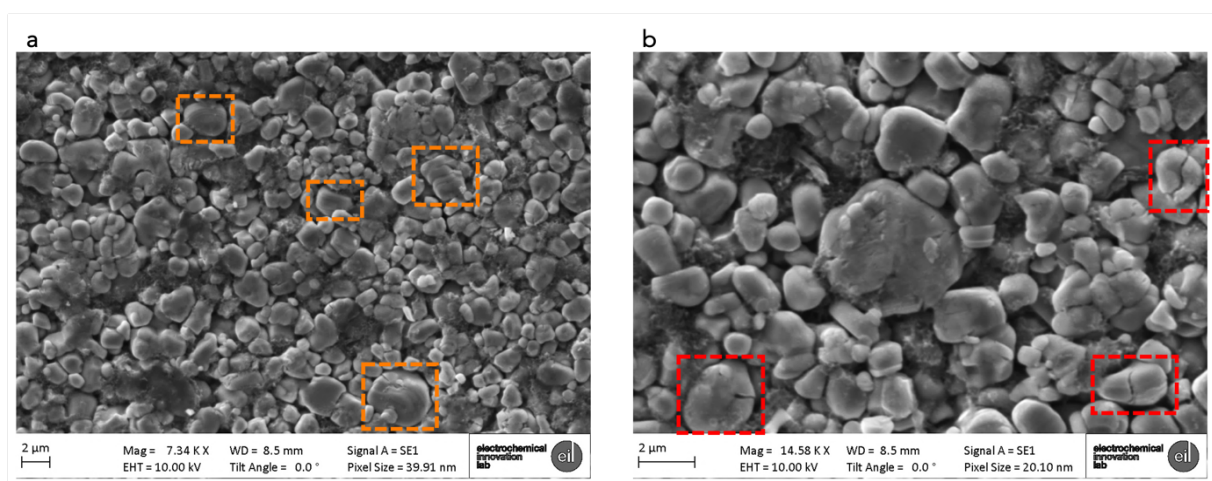

**Fig. S4:** Scanning electron microscopy images of the 500-cycle SC811 sample, showing several crystals with planar gliding defects in orange dashed boxes and single-crystal cracking in red dashed boxes.

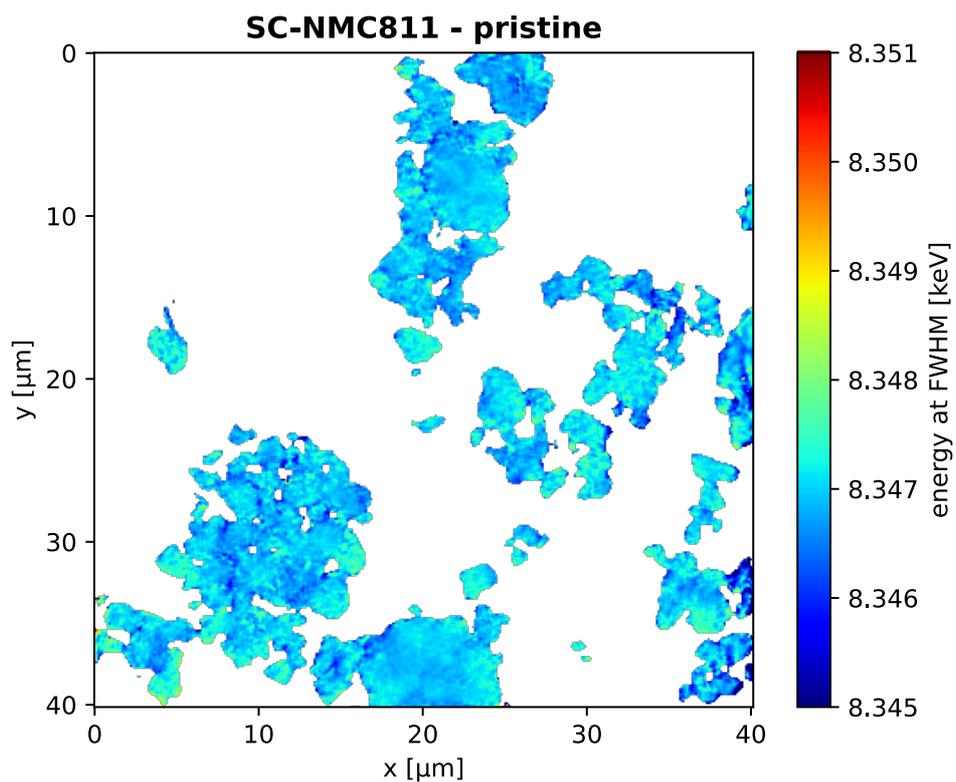

**Fig. S5:** 2D-Spectro-Ptychography from pristine SC-NMC811 sample over a 40 x 40 µm² area.

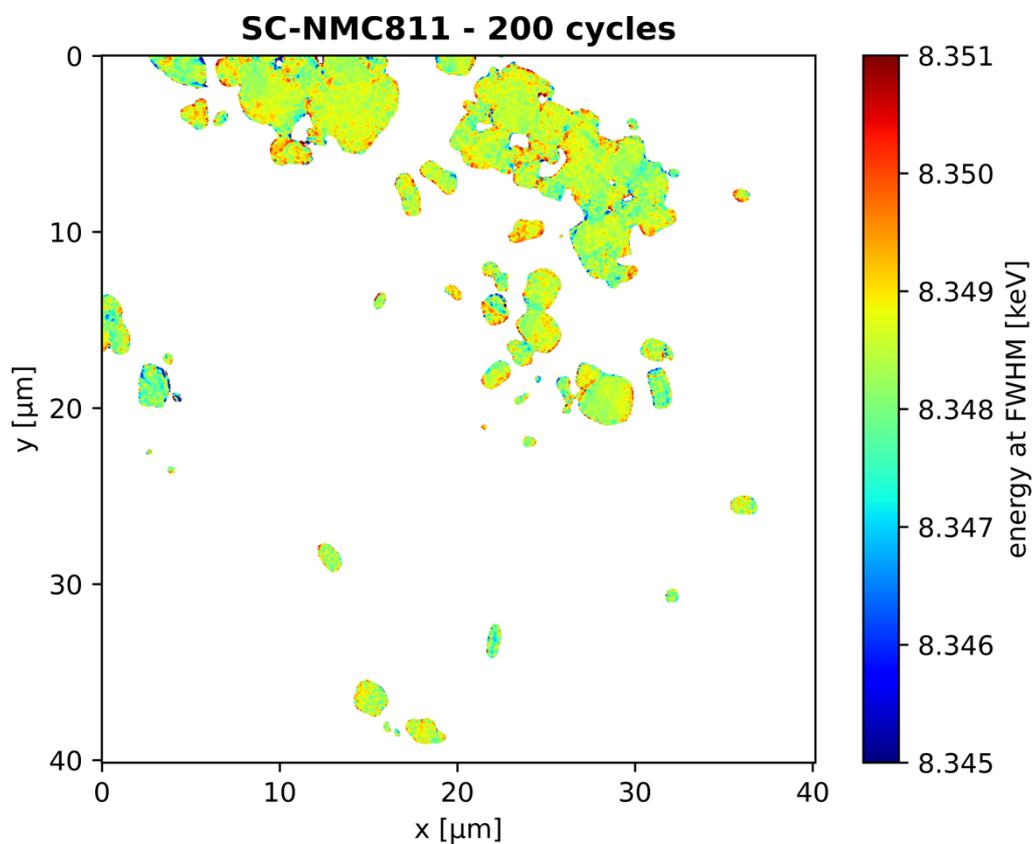

**Fig. S6:** 2D-Spectro-Ptychography from 200-cycle SC-NMC811, sample 1 over a 40 x 40 µm² area.

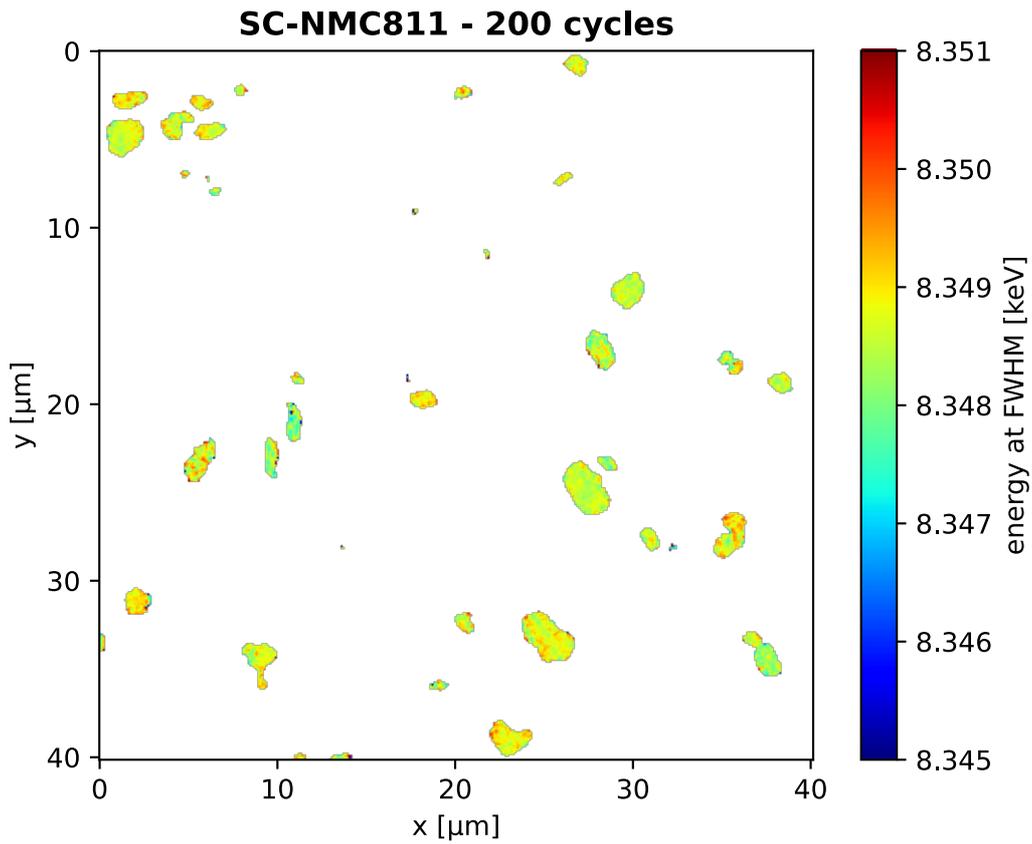
**Fig. S7:** 2D-Spectro-Ptychography from 200-cycle SC-NMC811, sample 2 over a 40 x 40 µm² area.

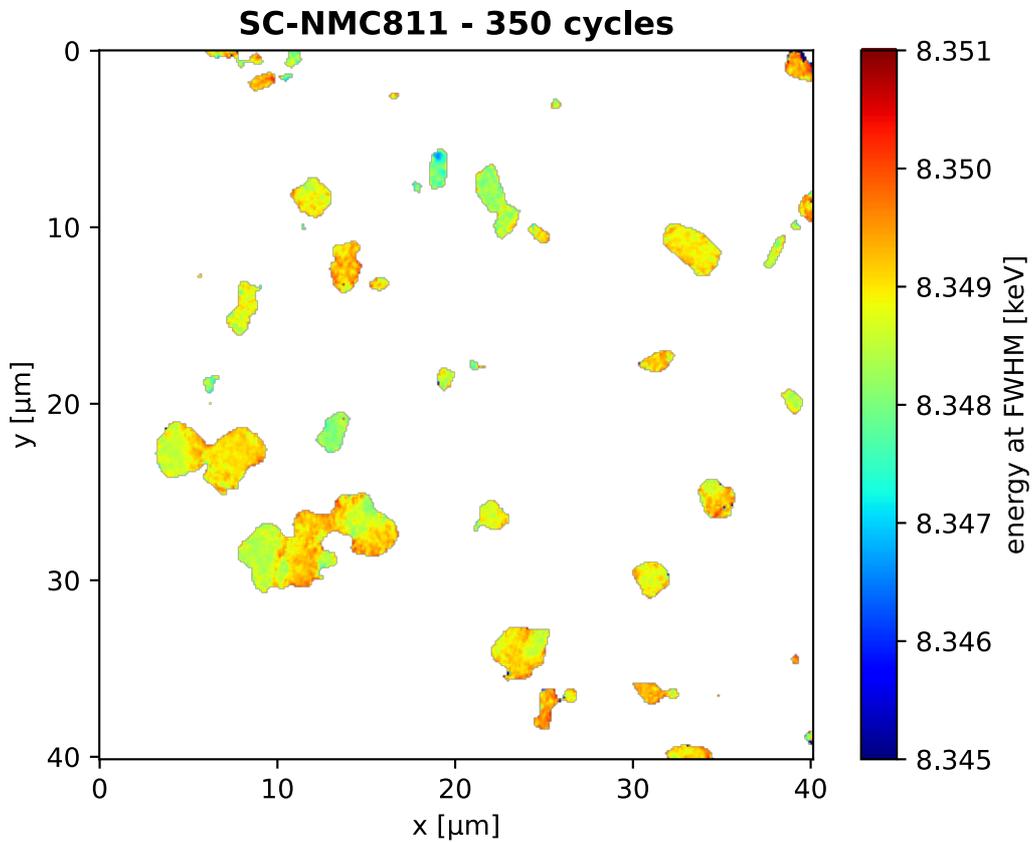
**Fig. S8:** 2D-Spectro-Ptychography from a 350-cycle SC-NMC811 sample over a 40 x 40 µm² area.

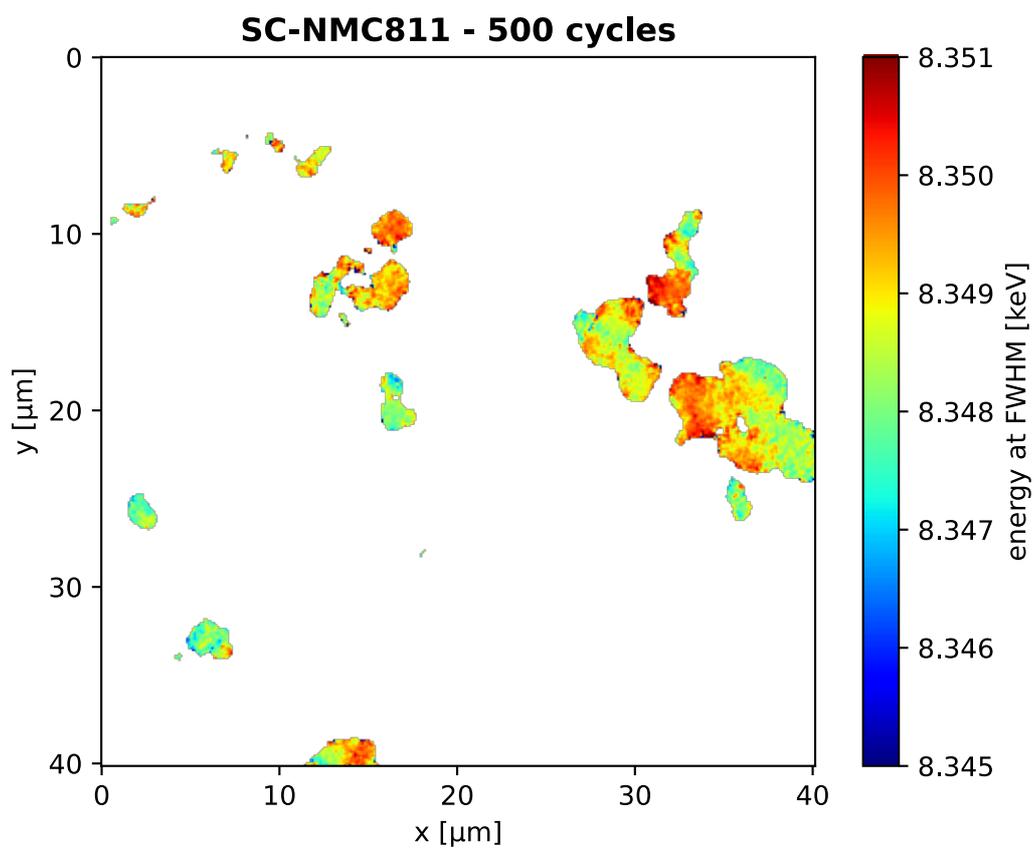

**Fig. S9:** 2D-Spectro-Ptychography from 500-cycle SC-NMC811, sample 1 over a 40 x 40 µm² area.

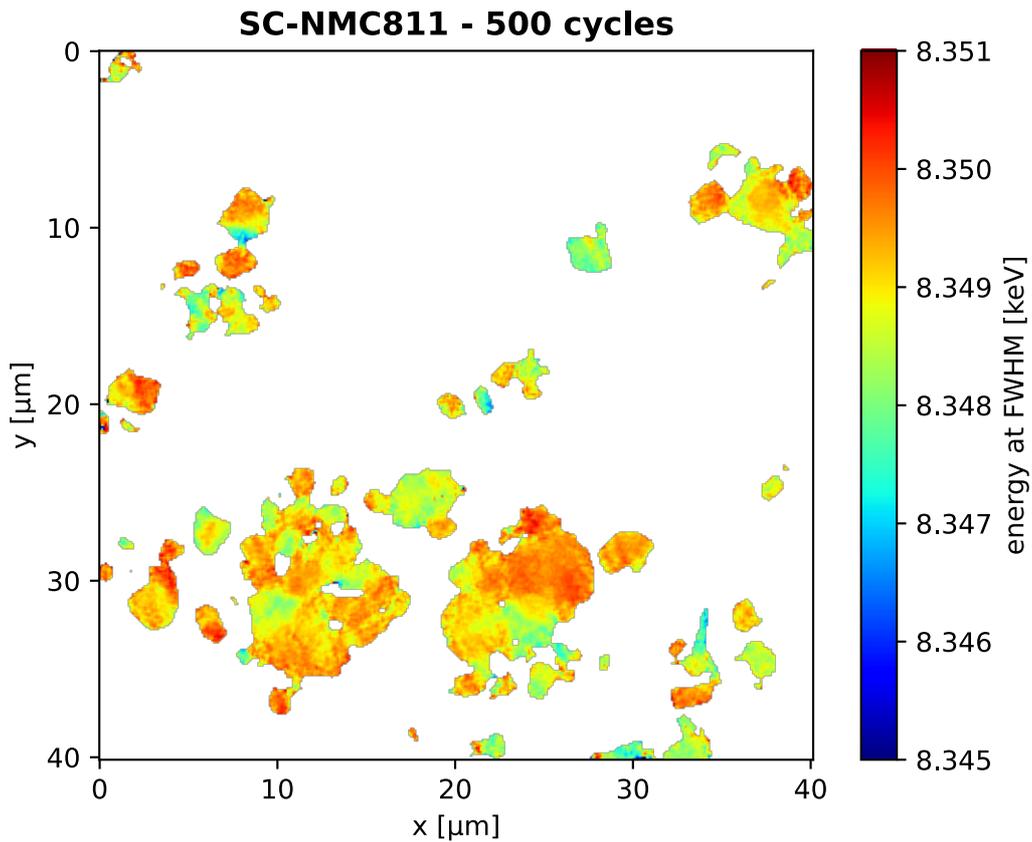

**Fig. S10:** 2D-Spectro-Ptychography from 500-cycle SC-NMC811, sample 2 over a 40 x 40 µm² area.

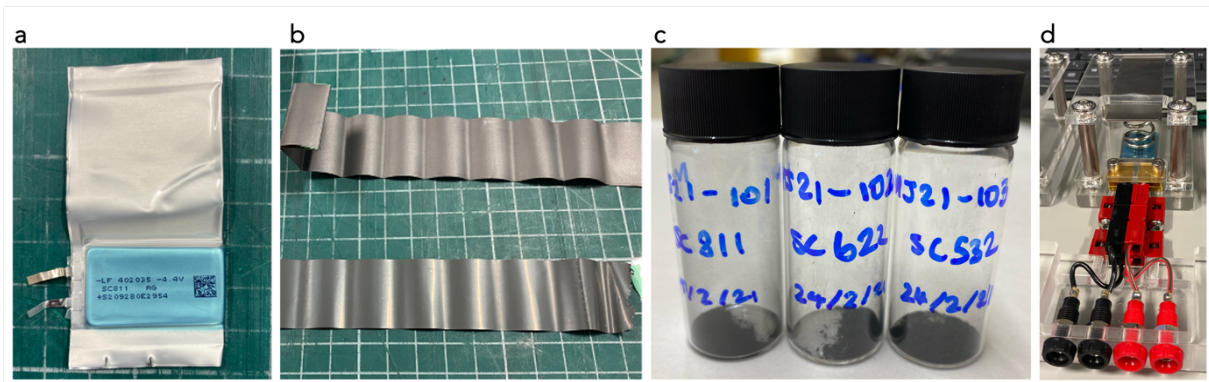

**Fig. S11: a,** Image of the commercially manufactured LiFUN cells used in these experiments. **b,** Images of the unravelled positive electrode (SC811 on aluminium foil) and negative electrode (artificial graphite on copper foil). **c,** Image of the harvested active material from the positive electrode. **d,** LiFUN pouch cell compression jig for lifetime cycling.

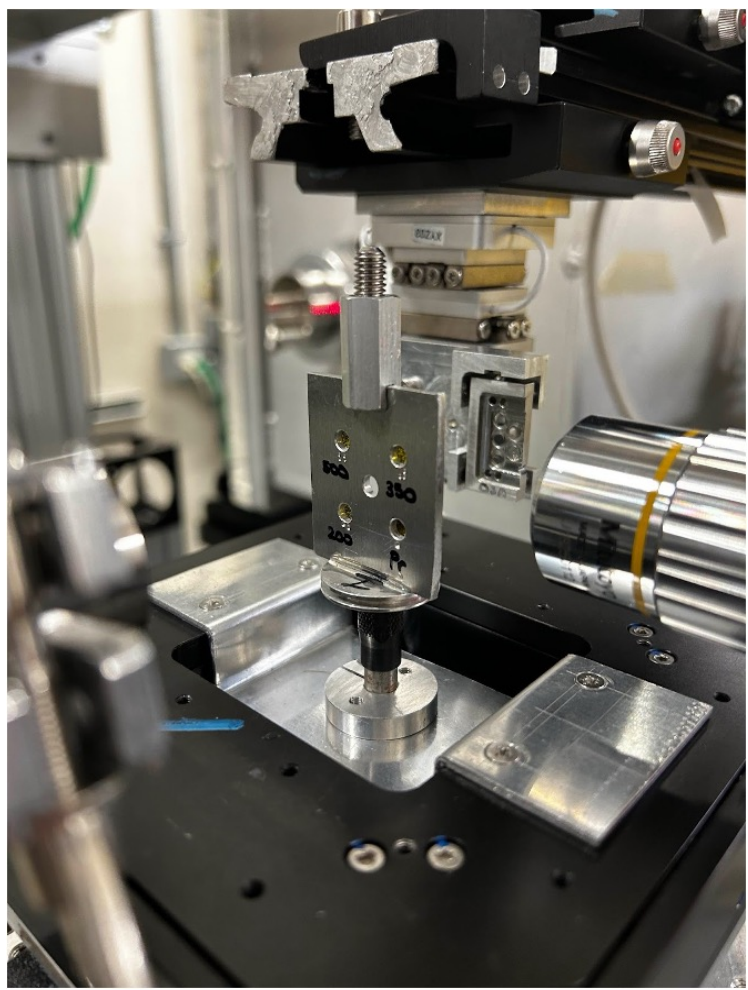

**Fig. S12:** Example of the Kapton grid used for the 2D-XANES-Ptychography experiments, with the pristine, 200, 350 and 500 samples clearly labelled.

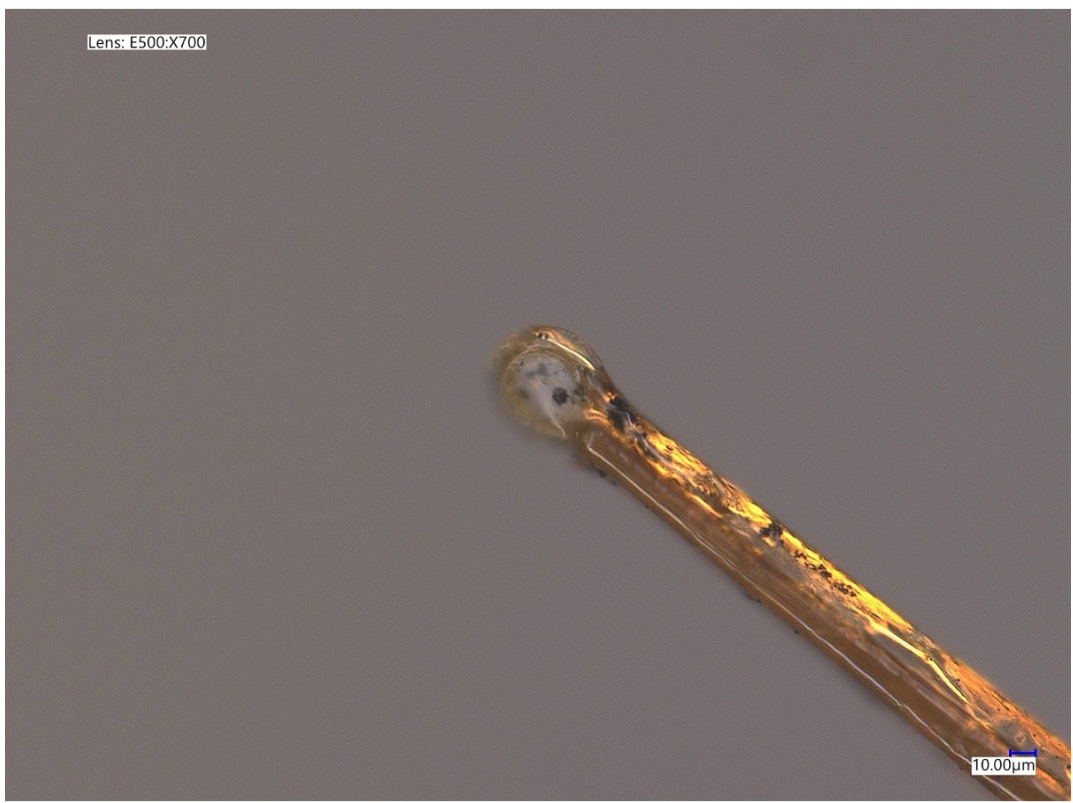

**Fig. S13:** Example of the Kapton loop crystal mount used for the 3D-XANES-Ptychography experiments, with the pristine sample shown in figure 2 of the main manuscript mounted.

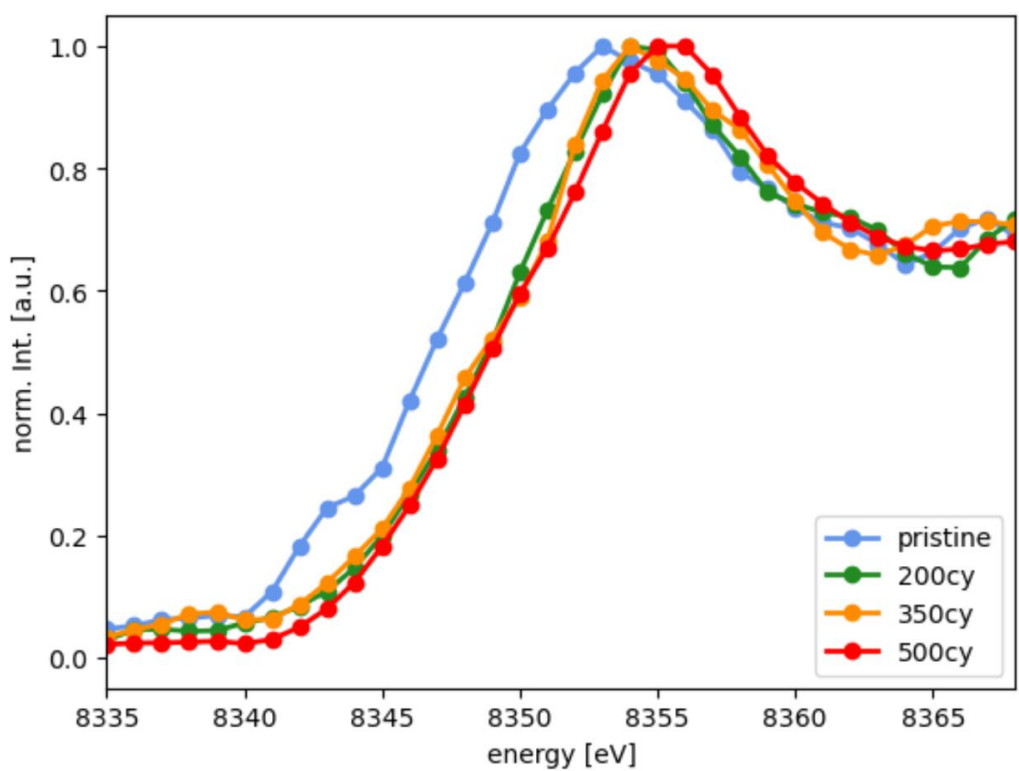

**Fig. S14:** Averaged XANES plots generated from the 2D-spectro-ptychography results for the SC-NMC811 at pristine, 200, 350 and 500 cycles over a 40 x 40 µm$^2$ area.

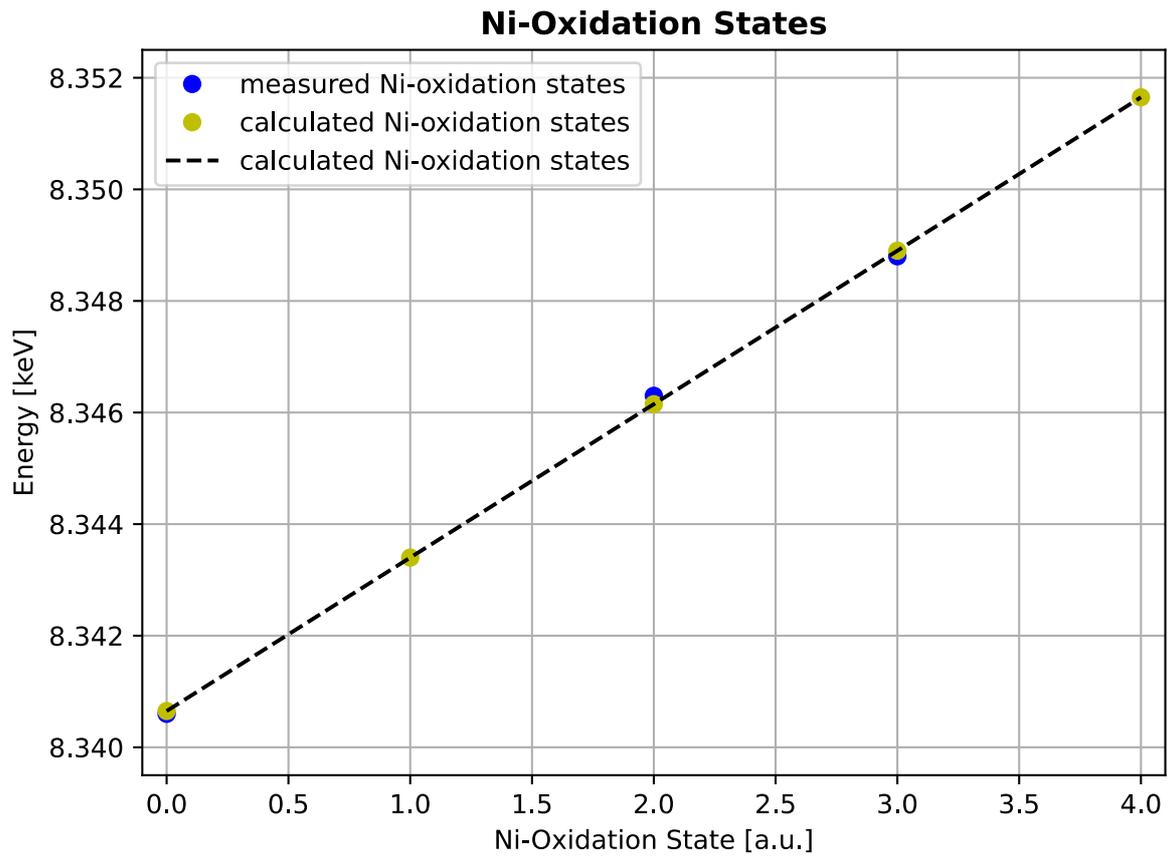

**Fig. S15:** Plot of the FWHM shifts in the XANES spectra vs. oxidation state of Nickel measured at the I13-1 instrument at Diamond Light Source.